# 平面全息阵列的超指向性研究

林航[1]　薛刘荀[1]　孙舒[1*]　高锐锋[2]　王珏[3]　王腾蛟[4]

（1. 上海交通大学信息与电子工程学院，上海 200240；2. 南通大学交通与土木工程学院，南通 226019；3. 南通大学信息科学技术学院，南通 226019；4. 华为无线网络研究部，上海 201700）

**摘　要**　本文研究了全息多输入多输出通信系统中均匀矩形阵列的超指向性。通过建立均匀矩形阵列的指向性数学模型，本文推导出了最大指向性的解析表达式，并结合数值仿真进行了系统性分析。研究结果表明，通过合理利用耦合效应可显著增强阵列指向性，但当天线间距远小于波长尺度时，这种增强效果会呈现边际递减现象。本研究为超指向性均匀矩形阵列的设计提供了理论基础，并为 5G/6G 通信系统中全息阵列的优化提供了重要参考。

**关键词**　全息 MIMO；互耦；超指向性；均匀矩形阵列；MIMO 无线通信

## Investigation of superdirectivity in planar holographic arrays

**LIN Hang[1]　XUE Liuxun[1]　SUN Shu[1]　GAO Ruifeng[2]　WANG Jue[3]　WANG Tengjiao[4]**

(1. School of Information Science and Electronic Engineering, Shanghai Jiao Tong University, Shanghai 200240, China; 2. School of Transportation and Civil Engineering, Nantong University, Nantong 226019, China; 3. School of Information Science and Technology, Nantong University, Nantong 226019, China; 4. Department of Wireless Network Research, Huawei Technologies CO., Ltd, Shanghai 201700, China)

**Abstract**　This paper studies the superdirectivity characteristics of uniform rectangular arrays (URAs) for holographic multiple-input multiple-output systems. By establishing a mathematical directivity model for the URA, an analytical expression for the maximum directivity is derived. Accordingly, systematic analysis is performed in conjunction with numerical simulations. Results show that the directivity can be significantly enhanced via rational utilization of coupling effects. However, this enhancement yields diminishing returns when antenna spacings transition to deep sub-wavelength scales. This study provides a theoretical basis for the design of superdirective URAs and offers valuable insights for holographic array optimization in 5G/6G communication systems.

**Keywords**　Holographic MIMO; mutual coupling; superdirectivity; uniform rectangular array; MIMO wireless communication

## 0　引言

全息多输入多输出（Holographic MIMO，HMIMO）是无线通信领域的一种前沿技术范式，它利用超密集天线阵列实现近乎连续的孔径表面[1-2]，实现了传统多输入多输出（multiple-input multiple-output, MIMO）系统的变革性突破。较小的天线间距导致天线之间产生较强的耦合效应，导致能量空间分布发生变化，可实现超过传统阵列的指向性增益[3-4]，称为超指向性。

HMIMO 系统的超指向性不仅有望显著提高无线通信系统的性能，也为无线电力传输等新兴应用提供了潜在的技术支持[3]。为了实现超指向性，[5]提出了一种新的耦合矩阵合成方法来确定超定向天线阵列的最优波束成形矢量，其中耦合矩阵可从球面波扩展法和有源分量方向图中推导。在此基础上，为了解决图案畸变并增强波束成形性能，[6]引入了一种利用三维超指向全息阵列的电磁混合波束成形方案，实现了相对均匀的超指向性增益，并实现了整个空间域的可编程波束聚焦。[7]的工作进一步证明，对于线性阵列中间距小于半波长的固定天线，随着天线数量的增加，可实现超指向性。

均匀矩形阵列（uniform rectangular array, URA）是传统 MIMO 中常见的阵列架构，通过二维布局实现多空间自由度和灵活的

三维波束成形，显著提高信道容量和覆盖范围。确定 URA 超指向性的现有方法主要依赖于计算密集型全波电磁仿真[5-6]，缺乏通用的分析框架。此外，URA 可实现的超指向性范围也值得研究，这对于验证仿真和评估实际应用至关重要。为此，本文对 URA 进行数学推导和理论研究。本文的主要贡献包括两方面：首先，给出了 URA 远场指向性的数学框架，并推导了解析表达式；其次，深入对比了常规和密集间距 URA 的指向性特性，系统分析了各种因素对指向性的影响。

## 1 系统模型

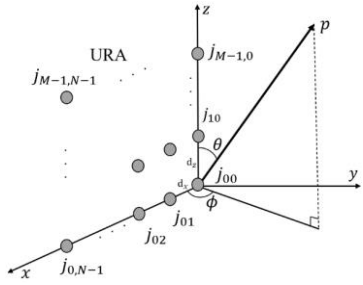

图 1  xz 平面上的 URA

考虑一个由 $M \times N$ 个单极化全向天线组成的 URA，如图 1 所示。天线间距沿 x 轴为 $d_x$，沿 z 轴为 $d_z$。天线由复数电流矢量

$$\mathbf{j} = (j_{00}, j_{01}, \ldots, j_{0,N-1}, j_{10}, \ldots, j_{M-1,N-1})^T \in \mathbb{C}^{MN}$$

激发。在图 1 中，$\mathbf{p}$ 表示传播向量，$\theta \in [0,\pi]$ 表示天顶角，$\phi \in [0,\pi]$ 表示方位角。天线沿 x 轴和 z 轴分布的离散空间频率可表示为：

$$f_x = \frac{d_x}{\lambda}(\mathbf{p} \cdot \mathbf{x}) = \frac{d_x}{\lambda}\sin\theta\cos\phi \quad (1)$$

$$f_z = \frac{d_z}{\lambda}(\mathbf{p} \cdot \hat{\mathbf{z}}) = \frac{d_z}{\lambda}\cos\theta \quad (2)$$

式中：$\mathbf{p}$ 表示传播路径方向的单位矢量；$\mathbf{x}$ 表示 x 轴方向的单位矢量；$\hat{\mathbf{z}}$ 表示 z 轴方向的单位矢量。给定电流向量，阵列在传播方向上的输出信号为

$$J\left(\phi,\theta,\frac{d_x}{\lambda},\frac{d_z}{\lambda}\right) = \mathbf{a}^H\left(\phi,\theta,\frac{d_x}{\lambda},\frac{d_z}{\lambda}\right)\mathbf{j}$$

$$= \sum_{m=0}^{M-1}\sum_{n=0}^{N-1} j_{mn} e^{-j2\pi f_x n} e^{-j2\pi f_z m} \quad (3)$$

$$\mathbf{a}\left(\phi,\theta,\frac{d_x}{\lambda},\frac{d_z}{\lambda}\right) = (1,\ldots,e^{j2\pi(N-1)f_x},$$

$$e^{j2\pi f_z},\ldots,e^{j2\pi(M-1)f_z + j2\pi(N-1)f_x})^T \quad (4)$$

式中：$\mathbf{a}\left(\phi,\theta,\dfrac{d_x}{\lambda},\dfrac{d_z}{\lambda}\right)$ 为阵列远场[8]导向矢量。从信号变换的角度来看，如果将电流向量视为空间采样的离散信号，则数学模型与二维离散傅里叶变换同构。

特定方向上的指向性或阵列增益 $(\phi,\theta)$ 定义为该方向的功率谱密度与其在空间 $[0,\pi] \times [0,\pi]$ 上的平均值之比

$$G\left(\mathbf{j},\phi,\theta,\frac{d_x}{\lambda},\frac{d_z}{\lambda}\right) =$$

$$\frac{2\pi\left|J\left(\phi,\theta,\dfrac{d_x}{\lambda},\dfrac{d_z}{\lambda}\right)\right|^2}{\int_0^\pi\int_0^\pi\left|J\left(\alpha,\beta,\dfrac{d_x}{\lambda},\dfrac{d_z}{\lambda}\right)\right|^2\sin\beta\,d\alpha\,d\beta} \quad (5)$$

式中因子 $2\pi$ 和 $\sin\beta$ 源于定义域从波数变为球形[9]。为简单起见，我们首先进行改写

$$\left|J\left(\phi,\theta,\frac{d_x}{\lambda},\frac{d_z}{\lambda}\right)\right|^2 = \mathbf{j}^H \cdot$$

$$\mathbf{a}\left(\phi,\theta,\frac{d_x}{\lambda},\frac{d_z}{\lambda}\right)\mathbf{a}^H\left(\phi,\theta,\frac{d_x}{\lambda},\frac{d_z}{\lambda}\right)\mathbf{j} \quad (6)$$

令 $\mathbf{A}(\phi,\theta,\dfrac{d_x}{\lambda},\dfrac{d_z}{\lambda}) = \mathbf{a}(\phi,\theta,\dfrac{d_x}{\lambda},\dfrac{d_z}{\lambda})\mathbf{a}^H(\phi,\theta,\dfrac{d_x}{\lambda},\dfrac{d_z}{\lambda})$。

为了进一步简化表达式，需要对式（5）中的分母进行详细计算，可利用以下积分性质

$$\int_0^\pi\int_0^\pi e^{j\pi u\sin\theta\cos\phi} e^{j\pi v\cos\theta}\sin\theta\,d\phi\,d\theta$$

$$= 2\pi\,\text{sinc}\left(\sqrt{u^2+v^2}\right) \quad (7)$$

式中 $\text{sinc}(x) = \frac{\sin(\pi x)}{\pi x}$。利用上述结论，式(5)可化简为

$$G\left(\mathbf{j}, \phi, \theta, \frac{d_x}{\lambda}, \frac{d_z}{\lambda}\right) = \frac{\mathbf{j}^H \mathbf{A}(\phi, \theta, \frac{d_x}{\lambda}, \frac{d_z}{\lambda})\mathbf{j}}{\mathbf{j}^H \mathbf{C}(\frac{d_x}{\lambda}, \frac{d_z}{\lambda})\mathbf{j}} \quad (8)$$

式中 $\mathbf{C}\left(\frac{d_x}{\lambda}, \frac{d_z}{\lambda}\right)$ 具体表示为

$$\left(\mathbf{C}\left(\frac{d_x}{\lambda}, \frac{d_z}{\lambda}\right)\right)_{i,j} = \text{sinc}\left(\frac{2}{\lambda}\sqrt{\begin{array}{l}d_x^2(\text{mod}(i-1,N) \\ -\text{mod}(j-1,N))^2 \\ +d_z^2\left(\left\lfloor\frac{i-1}{N}\right\rfloor - \left\lfloor\frac{j-1}{N}\right\rfloor\right)^2\end{array}}\right) \quad (9)$$

因此，对于给定的电流矢量 $\mathbf{j}$，URA 中的线间距 $(d_x, d_z)$ 决定了 $\mathbf{C}\left(\frac{d_x}{\lambda}, \frac{d_z}{\lambda}\right)$，而具体方向反映在 $\mathbf{A}(\phi, \theta, \frac{d_x}{\lambda}, \frac{d_z}{\lambda})$ 中。这些属性共同影响 URA 在特定方向上的指向性。

对于天线间距为 $(d_x, d_z)$ 的给定阵列和方向 $(\phi, \theta)$，我们的目标是找到一个电流矢量 $\mathbf{j}$，该矢量在指定方向上最大化阵列的指向性。这个优化问题可表述为

$$\max_{\mathbf{j}} \frac{\mathbf{j}^H \mathbf{A}(\phi, \theta, \frac{d_x}{\lambda}, \frac{d_z}{\lambda})\mathbf{j}}{\mathbf{j}^H \mathbf{C}(\frac{d_x}{\lambda}, \frac{d_z}{\lambda})\mathbf{j}} \quad (10)$$

该优化问题是一个广义瑞利商问题，可转化为广义特征值问题：

$$\mathbf{A}\left(\phi, \theta, \frac{d_x}{\lambda}, \frac{d_z}{\lambda}\right)\mathbf{v}_k = \lambda_k \mathbf{C}\left(\frac{d_x}{\lambda}, \frac{d_z}{\lambda}\right)\mathbf{v}_k \quad (11)$$
$$k = 0, 1, \ldots, MN-1$$

式中：$\lambda_k$ 和 $\mathbf{v}_k$ 是 $\mathbf{A}(\phi, \theta, \frac{d_x}{\lambda}, \frac{d_z}{\lambda})$ 和相关特征向量的第 $k$ 个有序（按降序大小排序）特征值。$\mathbf{A}(\phi, \theta, \frac{d_x}{\lambda}, \frac{d_z}{\lambda})$ 是一个秩为 1 的矩阵，这意味着 $\lambda_0$ 是其唯一的正特征值。注意到对于任何 $\mathbf{A} \in \mathbb{C}^{X \times Y}$ 和 $\mathbf{B} \in \mathbb{C}^{Y \times X}$，矩阵 $\mathbf{AB}$ 和 $\mathbf{BA}$ 的非零特征值是相同的。应用该结论，可确定

$$\lambda_0 = \mathbf{a}^H\left(\phi, \theta, \frac{d_x}{\lambda}, \frac{d_z}{\lambda}\right) \\ \cdot \mathbf{C}^{-1}\left(\frac{d_x}{\lambda}, \frac{d_z}{\lambda}\right)\mathbf{a}\left(\phi, \theta, \frac{d_x}{\lambda}, \frac{d_z}{\lambda}\right) \quad (12)$$

因此，最大指向性为

$$G^*\left(\phi, \theta, \frac{d_x}{\lambda}, \frac{d_z}{\lambda}\right) = \mathbf{a}^H\left(\phi, \theta, \frac{d_x}{\lambda}, \frac{d_z}{\lambda}\right) \cdot \\ \mathbf{C}^{-1}\left(\frac{d_x}{\lambda}, \frac{d_z}{\lambda}\right)\mathbf{a}\left(\phi, \theta, \frac{d_x}{\lambda}, \frac{d_z}{\lambda}\right) \quad (13)$$

该表达式体现了最大指向性、辐射方向和天线间距之间的相关性。天线的耦合效应在 $\mathbf{C}^{-1}\left(\frac{d_x}{\lambda}, \frac{d_z}{\lambda}\right)$ 中体现。$G^*\left(\phi, \theta, \frac{d_x}{\lambda}, \frac{d_z}{\lambda}\right)$ 在半空间上的平均值是一个常数，表明耦合在 URA 面向的半空间上重新分配了最大增益。即使对于具有半波长间距的 URA，天线之间的耦合效应也是不可忽略的，因为偏离水平和垂直对齐的天线违反了半波长整数倍数间距准则，从而引起了非零耦合[4]。

## 2 数值结果和分析

对于具有半波长天线间隔的传统 URA，我们设置 $M=4, N=8$，所得指向性如图 2 所示。忽略耦合的指向性参考值为 $10\log_{10}(MN) = 15.05\,\text{dB}$。图 2 表明，传统 URA 在侧射方向 $(\phi = \frac{\pi}{2}, \theta = \frac{\pi}{2})$ 上可达到 16.68 dB 的指向性值，在端射平面（指本文中 URA 所在的平面）的 $\frac{\pi}{4}$ 和 $\frac{3\pi}{4}$ 方向

$(\phi=0,\theta=\dfrac{\pi}{4},\dfrac{3\pi}{4})$ 可达到 $16.65\,\mathrm{dB}$ 的指向性值,均超过 $10\log_{10}(MN)$。与传统的均匀线性阵列(uniform linear array, ULA)[7]相比,传统 URA 的指向性已经表现出由相互耦合产生的非均匀特性。

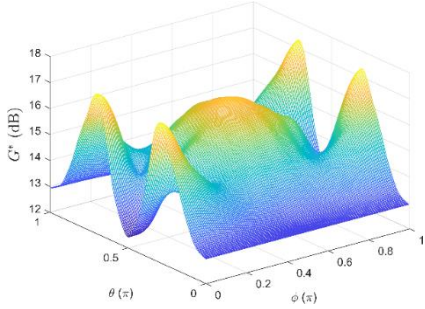

图 2　传统 URA 的指向性

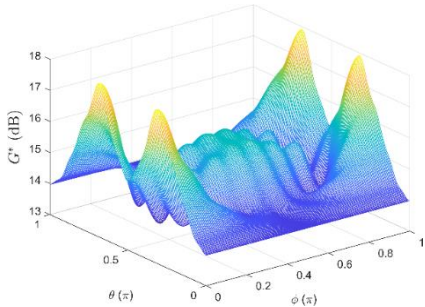

图 3　密集间距 URA 的指向性

密集间距 URA 的指向性如图 3 所示,图中参数设置为 $\dfrac{d_x}{\lambda}=\dfrac{d_z}{\lambda}=0.45, M=4, N=8$。随着阵列天线间距的减小,端射平面 $(\phi=0,\theta\in[0,\pi])$ 中的指向性显著增强,与半波长间距 URA 相比,最大值进一步增加,而侧射 $(\phi=\dfrac{\pi}{2},\theta=\dfrac{\pi}{2})$ 方向性减弱,这是由于最大指向性在空间中的重新分布。

为了探索天线间距如何影响指向性,图 4 描绘了具有不同天线间距的 URA 端射平面中的指向性,设定 $M=4, N=8$。当天线间距从 $0.5\lambda$ 减小到 $0.3\lambda$ 时,指向性显著增强,这归因于密集排列的天线之间的强耦合。然而,进一步将天线间距减小到 $0.1\lambda$,指向性在端射平面上仅表现出微小且大致均匀的增加。由于最小化天线间距可能会导致能效问题[7],因此确定天线间距阈值以在指向性和效率之间取得平衡至关重要。

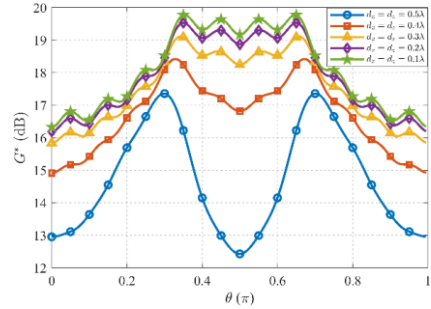

图 4　不同间距 URA 在端射平面上的指向性

## 3　结论

在本文中,我们深入研究了全息 URA 的指向性,包括传统配置和密集间距配置。我们推导了最大指向性的闭式表达式。分析表明,超指向性在端射平面中表现最为突出,峰值指向性与天线间距呈反关系。未来的工作可能会考虑包含实际约束的超指向性行为,例如天线辐射效率和非各向同性功率指向图,以及天线互耦对空间自由度和无线传输设计[10-11]等方面的影响。